\documentclass[12pt]{article}
\pdfoutput=1

\newlength{\abstractwidth}
\usepackage[nolabel, final]{showlabels} 

\usepackage{epsf}
\usepackage{color}
\usepackage{graphicx}
\usepackage{tikz}
\usepackage{hyperref}

\usepackage{amsmath}
\usepackage{amssymb}
\usepackage{latexsym}
\usepackage{showlabels}
\numberwithin{equation}{section}
\usepackage{mathtools}

\usepackage{tikz}
\usetikzlibrary{trees}
\usetikzlibrary{decorations.pathmorphing}
\usetikzlibrary{decorations.markings}

\usetikzlibrary{arrows}

\tikzset{
    photon/.style={decorate, decoration={snake}, draw=red},
    electron/.style={draw=blue, postaction={decorate},
        decoration={markings,mark=at position .55 with {\arrow[draw=blue]{>}}}},
    gluon/.style={decorate, draw=magenta,
        decoration={coil,amplitude=4pt, segment length=5pt}} 
}

\makeatletter
\renewcommand\section{\@startsection {section}{1}{\z@}%
                                   {-3.5ex \@plus -1ex \@minus -.2ex}
                                   {2.3ex \@plus.2ex}%
                                   {\normalfont\large\bfseries}}
\renewcommand\subsection{\@startsection{subsection}{2}{\z@}%
                                     {-3.25ex\@plus -1ex \@minus -.2ex}%
                                     {1.5ex \@plus .2ex}%
                                     {\normalfont\bfseries}}
\makeatother

\renewcommand{\thefootnote}{\fnsymbol{footnote}}
\renewcommand{\thanks}[1]{\footnote{#1}}
\newcommand{\starttext}{
\setcounter{footnote}{0}
\renewcommand{\thefootnote}{\arabic{footnote}}}

\newcommand{\bea}{\begin{eqnarray}}
\newcommand{\eea}{\end{eqnarray}}
\newcommand{\be}{\begin{eqnarray}}
\newcommand{\ee}{\end{eqnarray}}

\def\cA{{\cal A}}

\def\nn{\nonumber}

\def\CC{{\mathbb C}}

\def\CC{{\mathbb C}}

\def\Im{{\rm Im \,}}

\def\det{{\rm det \,}}

\begin{document}
\starttext
\setcounter{footnote}{0}

\begin{flushright}
QMUL-PH-19-21\
\end{flushright}

\vskip 0.3in

\begin{center}

{\Large \bf Superstring Amplitudes, Unitarity, and \\ \vskip 0.3cm Hankel  Determinants of Multiple Zeta Values}

\vskip 0.2in

{\large  Michael B. Green$^{(a,b)}$ and Congkao Wen$^{(a)}$}

\vskip 0.15in

{\tt \small  M.B.Green@damtp.cam.ac.uk,  c.wen@qmul.ac.uk}

\vskip 0.15in

{ \sl (a) Centre for Research in String Theory, School of Physics and Astronomy, }\\
{\sl Queen Mary University of London, Mile End Road, London, E1 4NS, UK}

\vskip 0.1in

{ \sl (b) Department of Applied Mathematics and Theoretical Physics, }\\
{\sl University of Cambridge, Wilberforce Road, Cambridge CB3 0WA, UK}

\vskip 0.2in

\begin{abstract}
\vskip 0.1in
  The interplay of unitarity and analyticity has long been known to impose strong constraints on scattering amplitudes in quantum field theory and string theory.  This has been highlighted in recent times  in a number of papers and lecture notes.  Here we examine such conditions in the context  of superstring tree-level scattering amplitudes, leading to positivity constraints on determinants  of Hankel matrices involving polynomials of multiple zeta values.  These  generalise certain constraints on polynomials of single zeta values in the mathematics literature.

\end{abstract}
\end{center}

\newpage
 
\setcounter{tocdepth}{2} 
\tableofcontents
\newpage

\baselineskip=15pt
\setcounter{equation}{0}
\setcounter{footnote}{0}

\section{Introduction and outline}
\label{overview}
\setcounter{equation}{0}

The S-matrix programme of the 1960s demonstrated that a number of very striking properties of scattering amplitudes arise from a few powerful principles, such as unitarity and analyticity.   A key insight that arose from the analysis of  hadronic experiments around 1967 is the idea of duality in hadronic processes -- the sum of an infinite number of resonance poles in the $s$-channel reproduces the sum of an infinite set of poles in the  $t$-channel.  This  is the ``dual resonance'' realisation of the bootstrap programme and was an essential feature of the Veneziano model \cite{Veneziano:1968yb}, which evolved into the bosonic open-string theory, and the Virasoro model \cite{Virasoro:1969me}, which evolved into the bosonic closed-string theory. It is also a feature of large-$N$ limit of $SU(N)$ QCD \cite{tHooft:1973alw}

In recent years there has been a revival of these ideas in the context of important developments in quantum field theories related to the Standard Model and quantum gravity, as well as their supersymmetric counterparts. Unitarity together with analyticity lead to nontrivial conditions on low-energy effective field theory that imply that Wilson coefficients should respect certain positivity bounds~\cite{Adams:2006sv}.   Such positivity bounds have many important applications ranging from ruling out phenomenological models~\cite{Adams:2006sv} to proving the $a$-theorem in four dimensions \cite{Komargodski:2011vj}, and to a better understanding of the weak gravity conjecture \cite{ArkaniHamed:2006dz} (see e.g. \cite{Cheung:2014ega, Andriolo:2018lvp, Hamada:2018dde, Bellazzini:2019xts}). The results of~\cite{Adams:2006sv} have recently been extended  to an infinite set of positivity conditions on these coefficients by Arkani-Hamed, Huang and Huang, as described in~\cite{EFThedron, ArkaniHamed:2018,Huang:2018}.\footnote{Different kinds of improved positivity bounds on low-energy coefficients of EFTs that generalise the results of~\cite{Adams:2006sv} were  explored in~\cite{deRham:2017avq, deRham:2017xox}} Recent applications of these ideas can be found in \cite{Chen:2019qvr}, which explores constraints on low-energy spectrum of quantum field theory when coupled to gravity, as well as implications for the weak gravity conjecture. These ideas have also been applied to correlation functions in conformal field theories~\cite{Arkani-Hamed:2018ign} (see also \cite{Sen:2019lec}). 

Some background material will be reviewed in section~\ref{review}, following closely the ideas in ~\cite{EFThedron, ArkaniHamed:2018,Huang:2018}. A general  consequence of unitary and analyticity in the Mandelstam invariants ($s,t,u$)  that follows from simple dispersion relations  is that coefficients in the low energy expansion of the four-particle amplitude are constrained by ultraviolet properties of the amplitude.  These coefficients can be assembled into Hankel matrices\footnote{ A Hankel matrix  is a $m\times n$ matrix  $H[a]_{ij}$  in which $a_{ij}$  depends only on $i+j$. For our purposes it is sufficient to  consider  square Hankel matrices for which $m=n$.} that must  generally be totally non-negative -- i.e. all  minors of such matrices are non-negative.   For theories that exhibit duality, such as string theory or large-$N$ QCD, which contain an infinite number of particle states with unbounded masses and spins, the coefficients of the Hankel matrices are  necessarily totally positive. These constraints follow from the fact that the low energy coefficients form a Stieltjes half-moment sequence \cite{independent-Hankel}.
  The precise statement of these constraints on the  low-energy coefficients  depends on properties of the ultraviolet behaviour of the amplitude.  Furthermore, there are particular subtleties in the general analysis of theories  with massless particles, which have massless threshold branch cuts in the complex $s$-plane and are also singular in the forward direction \cite{deRham:2017xox}.
 
 This paper explores these constraints on the low-energy expansion of  tree amplitudes describing the scattering of four massless particles in open and in closed  superstring theories that illustrate these points.  Since we restrict our considerations to tree amplitudes we will avoid dealing with the subtlety of massless thresholds.  Furthermore, we will see that there is no problem in subtracting the massless exchange contributions that are singular in the forward direction.
 
Clearly there is a limited amount of information that can be obtained by considering only massless four-particle tree amplitudes.    For example, in order to sample all the information contained in the no-ghost theorem in string theory \cite{Brower:1972wj,Goddard:1972iy,Thorn:1983cz} it would be necessary to consider amplitudes with arbitrary numbers of massless external scattering particles, or to consider all four-particle amplitudes with arbitrary massless and massive external states. 

Nevertheless the information in massless  four-article tree amplitudes has some direct connections with mathematical considerations.   As is  well-known, in the case of the open superstring the  coefficients in the low-energy expansion are   rational polynomials in Riemann zeta values.    As will be seen in section~\ref{sec:open}, this leads to positivity conditions on determinants of Hankel matrices of zeta values that extend those discussed in~\cite{Monien:2009,Haynes:2015}.   

In section~~\ref{sec:closed} we will see that the situation is more subtle in the closed-string case.   In this case the Hankel matrix constraints do not apply to the low-energy expansion of the full four-particle amplitude at fixed $t$ due to negative contributions from the $u$-channel poles.  In order to analyse the unitarity constraints  we will separate the fixed-$t$  amplitude into the sum of the contribution of $s$-channel poles and the contribution of $u$-channel poles.   The low energy expansion of the contribution from the $s$-channel poles contains a combination of irreducible Multiple Zeta Values (MZVs),\footnote{An {\it irreducible} MZV is one that cannot be expressed as a rational  polynomial in single  zeta values.} which cancel in the full four-particle amplitude (resulting in coefficients that are rational polynomials of odd zeta values).
We will argue that the unitarity conditions lead to total positivity of the Hankel matrices built out of these MZVs.     This gives a host of conditions on rational polynomials of particular combinations of MZVs.

These results are summarised and discussed in section~\ref{sec:discussion}.

\section{Unitarity constraints on low-energy expansion coefficients}
\label{review}

In this section we will follow the discussion in  \cite{EFThedron, ArkaniHamed:2018,Huang:2018} (see also~\cite{Chen:2019qvr}).
We begin by reviewing the general arguments that demonstrate some of the constraints of unitarity on four-particle scattering amplitudes.   If the scattering particles have equal masses, $\mu$, their momenta $k_r$ ($r=1,2,3,4$), are $D$-dimensional Minkowski vectors satisfying the mass-shell conditions   $k_r\cdot k_r=- \mu^2$.  Such amplitudes are  functions of the Mandelstam invariants $s_{ij} = -(k_i+k_j)^2 = 2\mu^2-2 k_i\cdot k_j$.  As usual we will use the notation $s_{12}=s_{34}=s$, $s_{14}=s_{23}=t$,  $s_{13}=s_{24}=u$, and recall that  momentum conservation $\sum_r k_r=0$ implies $s+t+u=4\mu^2$.  In addition, the external particles generally have non-zero spin and so the scattering data includes information about their polarisations, although this will be suppressed in most of the following.

 The amplitude $A_4(s,t)$  is an analytic function of $s$, $t$ and $u$,  apart from a very specific set of singularities.
The physical region for the  elastic scattering process $1+2\to 3+4$ with centre of mass energy  $2E=\sqrt s$ and scattering angle $\cos\theta =1+ 2t/(s-4\mu^2)$   is $s\ge  4\mu^2$, $4\mu^2-s\le t\le 0$, $4\mu^2-s\le u\le 0$.  The amplitudes for other physical regions in which $1+4\to 2+3$  and $1+3\to 2+4$ are related  by  appropriate analytic continuation (or crossing symmetry).
 The singularities of the amplitude include poles corresponding to intermediate bound states or resonances, normal threshold branch cuts corresponding to the production of intermediate multi-particle states, and anomalous thresholds that    lie outside the physical scattering region.\footnote{A basic review of the singularity structure of the S-matrix is given in 
\cite{Eden:1966dnq}.  However, this is restricted to the amplitudes with massive scattering particles with zero angular momenta.}  

We will be concerned with  theories that exhibit ``duality'', such as string theory, or the large-$N$ expansion of $SU(N)$ Yang--Mills theory.   The tree-level contribution to four-particle scattering amplitudes in such theories  (i.e. the leading perturbative contributions in string perturbation theory, or in the $1/N$ expansion in $SU(N)$ Yang--Mills theory)  necessarily possess an infinite sequence of poles at positions $m_a^2$ ($a=1,\dots,\infty$)  along  the positive real $s$, $t$ and $u$ axes.  In the full non-perturbative amplitude  unitarity implies the presence of branch cuts along the real $s$, $t$ and $u$ axes and almost all the poles are shifted below the real $s$, $t$ or $u$ axes and  are shielded  from the physical sheet behind branch cuts.  

A general feature of unitarity that plays a key r\^ole in the following is the optical theorem, which states
\begin{align}
\Im A_4(s,0) = \sqrt{s(s-4\mu^2})\, \sigma_{tot}(s) >0\,,
\label{optical}
\end{align}
where $\sigma_{tot}$ is the total cross section, which is positive.   A stronger set of positivity conditions of the form 
\begin{align}
\partial_t^n \, \Im A(s,t) \big|_{t=0} >0\,,\qquad \forall n>0\,
\label{strongplus}
\end{align}
can be deduced from properties of the partial wave expansion.

In this paper we will restrict attention to tree-level expressions in dual resonance  theories.  In such cases a scalar amplitude can be written in the  $s$-channel partial wave form\footnote{We are imposing the mass-shell condition $u=-s-t$ appropriate for scattering of massless states, in the following.}
\begin{align}
A_4(s,t) =A_4^{(s)}(s,t)+A_4^{(u)}(u,t)\,,
\label{fullamp}
\end{align}
where
\begin{align}
A^{(s)}_4
(s,t)=
 \sum _{a} \frac{p_{a} \, G_{\ell_a}^{\frac{D-3}{2}}(\cos\theta_a) }{m_a^2-s}\,, \qquad A^{(u)}_4 (u,t)= \sum _{a}  \frac{p_{a} \, G_{\ell_a}^{\frac{D-3}{2}}(\cos\theta_a) }{m_a^2-u}\, ,
\label{genamp}
\end{align}
and
\begin{align}
\cos \theta_a = 1 + {2t \over m_a^2}\,.
\label{xdef}
\end{align}

These expressions are  sums of poles describing intermediate states with masses $m_a$ and angular momenta $\ell_a$ in the $s$ and $u$ channels.  
The residue of a given pole of angiular momentum $\ell$ is proportional to the Gegenbauer polynomial  $G_\ell^{\frac{D-3}{2}}(\cos\theta_a)$ (which is equal to the Legendre polynomial $P_\ell(\cos\theta_a)$ when $D=4$).\footnote{A generating function for Gegenbauer polynomials  may be defined by 
\begin{align}
\frac{1}{(1-2xt+t^2)^m}=\sum_{\ell=0}^\infty G_\ell^{(m)} (x)\, t^\ell   \, .
\label{gegendef}
\end{align}
}  The $s$-channel and $u$-channel centre of mass scattering angles are evaluated at the poles, so $\cos\theta _a= 1+2t/m_a^2$.
The proportionality constants $p_{a}$ denote the squares of the coupling between a pair of external scalar states and the intermediate  states of masses $m_a$ and angular momenta $\ell_a$, and so they are positive for a unitary theory.  Sincer  
$\partial_y^n \, G^{\frac{D-3}{2}}_{\ell}(y)\big |_{y=1}>0$  for all $n$,  the  positivity conditions \eqref{strongplus} are satisfied if the couplings satisfy  $p_a>0$ for all $a$.

The expression \eqref{genamp} is appropriate for describing the physical amplitude $1+2\to 3+4$ with $s>0$ and fixed $-s\le t\le0$.  However, it is defined  for all values of $s$, $t$, $u$ (with $s+t+u=0$) by analytic continuation.  The presence of  poles at $t\ge 0$ that contribute to the crossed process $1+4\to 2+3$ requires the index $a$ in \eqref{genamp} to take an infinite number of values so there is an infinite number of $s$-channel and/or $u$-channel poles with unbounded masses,  $m_a{\underset{a\to \infty}\to}\infty$.  Indeed, the requirement of an infinite number of tree-level poles is the hallmark of dual resonance models and  the gauge-singlet sectors of large-$N$ QCD.   

The structure of $A_4(s,t)$  differs  from that of a tree contribution to a conventional quantum field theory with a finite number of fields, in which there would be  explicit pole contributions in the $t$ channel, as well as in the $s$, and $u$  channels.  Adding $t$-channel poles would lead to an additional polynomial in $s$ in  \eqref{fullamp}, which would  markedly affect its large-$|s|$ behaviour.

More generally, the external states may have spin and this will be reflected by spin-dependent factors that complicate these expressions.  For the purpose of this paper, we will focus on the amplitudes where external states are scalars,  This is sufficient for our later considerations of maximally supersymmetric superstring amplitudes since in such theories a supermultiplet of massless states is described in terms of a Lorentz-scalar on-shell superfield.

An important technical point is that the amplitude is generally singular at the boundaries of the physical region when $s>0$ and  $t=0$ or $t=-s$, i.e. when $\theta=0$ or $\theta=\pi$.   In Yang--Mills theory and gravity this is due to the exchange of a massless gauge boson or graviton.  In order to discuss the low-energy expansion of the amplitude it is therefore important to subtract the singular term, which  we denote $A_{sing}(s,t)$.  Thus we may define a subtracted amplitude
\begin{align}
\tilde A_4 (s,t)= A_4(s,t)-A_{sing}(s,t) \, ,
\end{align}
 which is finite in the forward and backward limits, $t=0$ and $t=-s$.  
 
 The low energy expansion of $\tilde A_4(s,t)$  in powers of $s$ and $t$ is given by
\begin{align}
\tilde A_4(s,t) \big|_{s/m^2,t/m^2 \ll 1} = \sum_{p, q=0} g_{p,q}\, s^pt^q \, .
\label{double}
\end{align}
The coefficients in this expansion are  related to ultraviolet physics by virtue of the analyticity of the S-matrix by considering the Cauchy integral around a small circle enclosing the point $s=0$. For a fixed $t$, 
\begin{align}
\sum_q g_{p,q} t^q  &= \frac{1}{2\pi i} \oint \frac{ds}{s^{p+1} }\, \tilde A_4(s,t) \nn\\
&= \frac{1}{\pi} \int_{s=0}^\infty \frac{ds}{s^{p+1}} \Im \tilde A_4^{(s)}(s,t) +  \frac{1}{\pi} \int_{s=-t}^\infty \frac{ds}{s^{p+1}} 
\Im \tilde A_4^{(u)}(-s-t,t)  \nn\\
& +  \frac{1}{2\pi i}  \int_0^{2\pi} d\varphi \frac{ds}{s^{p+1} } \tilde A_4(s,t)\big|_{|s|\to \infty} \,,
 \label{cauchy}
\end{align}
where the integration contour in the first line   is a circle around the origin.   In passing to the second line the contour has been deformed  to pick up the contribution from the discontinuity  ${\rm Disc}\, A(s,t) =  \lim_{\epsilon \to 0} (A(s+i\epsilon ,t) -A(s-i\epsilon))  = 2i\Im \tilde A(s,t)  $.  
 If the subtracted amplitude behaves as $\tilde A_4(s,t) <| s|^w$ as  $s\to \infty$ the contribution from the contour at infinity (the third line in \eqref{cauchy})  vanishes for all $p\ge w$ (where we have defined $s= |s|\, e^{i \varphi}$).  From hereon we will assume that $w<0$ so this contribution can be dropped.  This will be shown to be true in the explicit examples that we will consider later although it is not true in most conventional field theory examples.

In the tree-level examples of the form \eqref{fullamp} $\Im \tilde A^{(s)}_4(s,t)$ is a sum of delta functions of the form\footnote{Note that the discontinuity of a pole at the origin in the complex $z$-plane  is given by ${\rm Disc} 1/z=-2 \pi i \delta(|z|)$.}
\begin{align}
 \Im \tilde A^{(s)}_4(s,t) =\pi  \sum_a p_a G_{\ell_a}^{\frac{D-3}{2}}\left(1+\frac{2t}{m_a^2}\right)     \,  \delta(s-m_a^2)\, ,
 \label{deltas}
\end{align}
 and so the contribution from the $s$-channel poles in \eqref{cauchy} is given by 
\begin{align}
\sum_q g^{(s)}_{p,q} t^q  =  \frac{1}{\pi} \int_{s=0}^\infty \frac{ds}{s^{p+1}} \Im \tilde A_4^{(s)}(s,t) =\sum_a   \frac{p_a}{m_a^{2p+2} } \, G_{\ell_a}^{\frac{D-2}{2}}  \left( 1+\frac{2t}{m_a^2}  \right)   \,.
 \label{scont}
\end{align}
It is important to note that \eqref{deltas} implies that  $\partial_t^n \Im \tilde A_4^{(s)}(s,t) |_{t=0} >0$.
 
 By contrast, the contribution of the $u$-channel poles is 
\begin{align}
\sum_q g^{(u)}_{p,q} t^q  = \frac{1}{\pi} \int_{s=-t}^{-\infty} \frac{ds}{s^{p+1}} \Im \tilde A_4^{(u)}(s,t) = \sum_a \frac{ (-1)^{p} p_a}   {(t+ m_a^2)^{p+1} }  \, G_{\ell_a}^{\frac{D-2}{2}}  \left( 1+\frac{2t}{m_a^2}  \right) \,.
 \label{ucont}
\end{align}
 It follows that $\partial_t^n \Im \tilde A_4^{(u))}(s,t) |_{t=0}$ has indefinite sign as do the coefficients $ g^{(u)}_{p,q} $.
The full coefficient of $s^pt^q$ is given by $g_{p,q}= g^{(s)}_{p,q} + g^{(u)}_{p,q}$ and therefore in general it has indefinite sign when there are $u$-channel poles, but in special cases $g_{p,q}$ is positive.

\subsection{The Stieltjes half-moment sequence}
It is useful to consider the $n\times n$ Hankel matrices  composed of the expansion coefficients $g_{p,q}$.
Thus, we may define a $(n+1)\times (n+1)$ Hankel matrix, which depends on $n+1$  sequential coefficients for a given value of $q$, ${\bf g_q}=\{g_{m,q},g_{m+1,q}\dots, g_{m+n,q}\}$
\begin{align}
 H_{n}^{(m)}[{\bf g_q}]  = \begin {pmatrix}
 g_{m,q} & g_{m+1,q}  & \cdots & g_{m+n,q} \\
 g_{m +1,q} & g_{m+2,q}  & \dots &  g_{m+ n+1,q}\\
  \vdots& \vdots &\vdots &\vdots  \\
 g_{m+n,q} & g_{m+n+1,q}  & \cdots& g_{m+2n,q}
 \cr
 \end{pmatrix} \, .
 \label{hnakdef}
\end{align}

The following theorem concerning positivity conditions on Hankel matrices  (an abbreviated version of  Theorem 2.8 in \cite{independent-Hankel}) is of central importance in the following:

{\it Given a sequence of real coefficients, ${\bf a}=(a_0, a_1, \dots, a_{\infty}$) then  the following statements are equivalent.
\begin{itemize}
\item
The infinite dimension Hankel matrix of the coefficients, $H_\infty^{(0)}({\bf a})$ is totally positive, so all its $n\times n$  minors are positive  for all $n$.
\item
The leading principal minors $\det H_{n}^{(0)} ({\bf{a})}$ and  $\det H_{n}^{(1)}({\bf{a})}$ are positive definite for all $n$.
\item
There is a positive measure $\mu$ on $[0,\infty)$ whose support is the infinite set  
\bea
a_k=\int_0^\infty y^k\, d\mu(y), \qquad \forall k\ge 0\,.
\label{moments}
\eea
This means that $(a_0, a_1, \dots, a_{\infty})$ form a Stieltjes half-moment sequence.  
\end{itemize}}

This theorem is relevant  if, for example,  we make the identifications $a_p \to g_{p,q}$, $y\to 1/s$ and $d\mu(y) \to  \partial_t^q \Im A^{(s)}(s,t)\big |_{t=0}\, ds/s$, where  $\Im A^{(s)}(s,t)$ is given in \eqref{deltas}.   

\subsection{Positivity conditions on moments of the amplitude}

We will now consider the first few terms in the expansion  of  $\tilde A_4(s,t)$ in powers of $t$, which leads to expressions for the individual entries of the matrix $g_{p,q}$.
\bigskip

{\it Terms of order $t^0$.}
\smallskip

The terms of lowest order in $t$ are the coefficients of $t^0$, which form the vector $g_{p,0}$ that is associated with the forward limit $t=0$ in \eqref{cauchy}.  In this case we have $g^{(s)}_{p,0} =g^{(u)}_{p,0}$, which results in 
\bea
g_{p,0} =\sum_a   \frac{2p_a}{m_a^{2p+2} }\, G_{\ell_a}^{\frac{D-2}{2}}  (1) \, ,
 \label{contodef}
\eea
 which is equivalent to 
 \bea
 \begin {pmatrix}
 g_{0,0}\\
 g_{1,0} \\
 g_{2,0}\\
 \vdots
  \cr
 \end{pmatrix}
 =\sum_a \frac{ 2p_a}{m_a^2}   G_{\ell_a}^{\frac{D-2}{2}}  (1) \, 
  \begin {pmatrix}
 x_a\\
 x_a^2\\
 x_a^3\\
\vdots
 \cr
 \end{pmatrix}
 \,,
 \label{matrixg}
 \eea
 where $x_a =m_a^{-2} \in R^+$.  
 The set of coefficients, ${\bf g_0}=\{g_{0,0},g_{1,0}\dots, g_{n,0}\}$ (for any $n$) defines a $(n+1)\times (n+1)$ Hankel matrix
 \be
 H_{n}^{(0)}[{\bf g_0}]= \begin {pmatrix}
 g_{0,0} & g_{1,0}  & g_{2,0} & \cdots & g_{n,0} \\
 g_{1,0} & g_{2,0} & g_{3,0} & \dots &  g_{ n+1,0}\\
 g_{2,0} & g_{3,0} & g_{4,0} &\cdots  & g_{n+2,0}  \\
  \vdots& \vdots &\vdots &\vdots & \vdots \\
 g_{n,0} & g_{n+1,0} &g_{n+2,0} & \cdots& g_{2n,0}
 \cr
 \end{pmatrix} \, .
 \label{hankdef}
 \eea

In this case we identify  $a_p=g_{p,0}$ in \eqref{moments} and the measure has the form $ \sum_a  2p_a\,  G_{\ell_a}^{\frac{D-2}{2}}  (1) /m_a^2\, \delta(y-x_a) dy$.  This is equivalent to the statement that the coefficients $g_{p,0}$  in \eqref{matrixg}  reside in the convex hull of points of a half moment curve with components $x_a^n$.
According to the theorem stated above  this property implies that the Hankel matrices built out of the coefficients $g_{m,0}$ are totally positive so all the minors of the Hankel matrix are positive definite.\footnote{Notational comments: A {\it minor} of a $n\times n$ matrix is the determinant of a $k\times k$ sub-matrix with $k< n$ obtained by deleting $n-k$ rows and $n-k$ columns.  A {\it principal minor} is the determinant of a $k\times k$ sub-matrix obtained by deleting $n-k$ columns and the $n-k$ rows that have the same numbering.  A {\it leading principal minor} is the determinant of a sub-matrix obtained by deleting the last $n-k$ columns and rows.}   

Using the expressions for $g_{m,0}$ in \eqref{matrixg} it is straightforward to show explicitly  that 
 \begin{align}
 \det  H_{n}^{(0)}[{\bf g_0}] &= \sum_P \prod_{i\in P} \frac{ p_i}{m_i^2} \prod_{i<j \in P}(x_i-x_j)^2 \geq 0\,, 
 \label{detform}
 \end{align}
 where $P$ is any length $n$  subset of  $\{1,2,\dots,L \}$,  with $L $ being the upper limit of the sum in \eqref{matrixg}. If $L $ is finite then it follows that $\det H_{n}^{(0)}[{\bf g_0}] =0$ for $n>L $, and $H_{n}^{(0)}[{\bf g_0}] $ is non-negative, i.e, $\det  H_{n}^{(0)}[{\bf g_0}]  \ge 0$. In  theories such as  string theory or large-$N$ QCD  the number of tree-level poles is infinite ($L=\infty$) and $ H_{\infty}^{(0)}[{\bf g_0}]$ is totally positive.  Analogously, one can show the positivity of $\det H_{n}^{(0)}[{\bf g_0}] $ with ${\bf g_0}=\{g_{m,0},g_{m+1,0}\dots, g_{m+n,0}\}$. The result is simply given by \eqref{detform} with $ \frac{ p_i}{m_i^2} \rightarrow \frac{ p_i}{m_i^2} x_i^m$. As stated previously,  the positivity of $\det H_{n}^{(0)}[{\bf g_0}]$ and $\det H_{n}^{(1)}[{\bf g_0}]$ is equivalent to the fact that the Hankel matrix $H_{\infty}^{(0)}[{\bf g_0}]$ is  totally positive.

\bigskip
{\it Terms of higher order in $t$.}
\smallskip

In order to consider the  low energy expansion of tree amplitudes to higher powers of $t$ we first expand the propagators in \eqref{fullamp} and \eqref{genamp} to all orders in $s$, giving 
\bea
 A^{(s)}_4(s,t) &=& \sum _a  \frac{ p_{a} }{m_a^2}\, G_{\ell_a}^{\frac{D-3}{2}} \left(1+\frac{2t}{m_a^2} \right)
 \sum^{\infty}_{n=0}   {s^{n}   \over m_a^{2n}}  \, ,
 \label{ampexp1}\\
 A^{(u)}_4(s,t)  &=& \sum _{a}\frac{ p_{a}}{m_a^2}\, G_{\ell_a}^{\frac{D-3}{2}}\left(1+\frac{2t}{m_a^2} \right)
 \sum^{\infty}_{n=0}   {(-s-t)^{n}   \over m_a^{2n}} \,.
\label{ampexp2}
\eea
Notice that the terms of order $s^{2n+1}\, t^0$ cancel in the sum of  $A_4^{(s)}(s,t)$ and $A_4^{(u)}(s,t)$, whereas the terms of order $s^{2n}\, t^0$ double,  giving the positive definite coefficients discussed earlier.
However, the  expansion of $A_4^{(u)}(s,t)$ in \eqref{ampexp2}  in powers of $s$ for a fixed power $t^q$ with $q>0$ has negative contributions that do not cancel with the terms in $A_4^{(s)}(s,t)$. Therefore the positivity condition for $A_4(s,t)$ is more involved, although more subtle positivity statements can still be obtained as in~\cite{EFThedron, ArkaniHamed:2018, Huang:2018} and \cite{deRham:2017avq, deRham:2017xox}.   For our purposes it will be sufficient to  consider properties of the low-energy expansions of  $A_4^{(s)}(s,t)$   and  $A_4^{(u)}(s,t)$ contributions separately.  

In the case of open-string amplitudes or large-$N$ QCD meson amplitudes the building blocks (such as the colour-stripped amplitudes)  only have $s$-channel poles (or $u$-channel poles) for a fixed value of $t$. 
In such cases only $A_4^{(s)} (s,t)$ (or $A_4^{(u)}(u,t)$) contributes to the low-energy expansion, and has an expansion of the form
\begin{align}
A^{(s)}_4(s,t)  &=\sum_{p,q=0}^\infty  \sum _{a}p_a\,   \frac{ 2^q} {q!\,m_a^{2}} \, \partial^q_{y} G_{\ell_a}^{\frac{D-3}{2}} \left(y \right){\big |}_{y=1}  \,
\left(   {s^{p} \over m_a^{2p}}  \right)\left(\frac{ t^q}{m_a^{2q}}\right) \, .
\label{posexpand}
\end{align}
Noting the property of Gegenbauer polynomials  $\partial^q_{y} G_{\ell_a}^{\frac{D-3}{2}} \left(y \right){\big |}_{y=1}>0$ and recalling that  $p_a>0$ the coefficients in \eqref{posexpand} satisfy 
\bea
 \frac{2^q\, p_a }{m_a^{2+2q}} \, \partial^q_{y} G_{\ell,a}^{\frac{D-3}{2}} \left(y \right){\big |}_{y=1}> 0\,.
 \label{poscoeffs}
 \eea    
So we see that, at any given order $t^q$, the low-energy expansion again  defines a half moment curve, so the Hankel matrices formed by the low-energy coefficients are totally non-negative. 

More generally,  amplitudes have both $s$-channel and $u$-channel poles so  both  $A^{(s)}_4(s,t)$  and $A^{(u)}_4(s,t)$ contribute.    In such cases the coefficients in the low energy expansion are not necessarily positive and do not reside on a moment curve (apart from the special $t^0$ case)  and there is no straightforward condition on the Hankel  matrices. However the coefficients of the low-energy expansion of the term with $s$-channel poles, $A_4^{(s)}(s,t)$ do satisfy positivity conditions that again generally lead to totally non-negative Hankel matrices.  As before, when the range of $a$ is infinite (as is the case with closed-string amplitudes and glueball amplitudes in large-$N$ QCD) the Hankel matrices are totally positive.  The term with $u$-channel poles, $A_4^{(u)}(u,t)$, satisfies the same conditions when expanded in powers of $u$ for a fixed power of $t$.  
We will see that in the closed-string case this leads to totally positive Hankel matrix determinants  with entries that are rational linear combinations of MZVs.   

 \section{Four-particle open superstring tree  amplitudes}  
 \label{sec:open}

In the rest of this paper we will consider critical superstring theory amplitudes, which have massless sectors describing maximally supersymmetric Yang--Mills theory in the case of open strings and maximally supersymmetric gravity in the case of closed strings.
 
 After stripping off the colour factors the amplitude that describes the scattering of any four massless particles in the Yang--Mills supermultiplet  has a term of the form 
  \bea
\cA_{op}(s,t) =  P_4\,  A_{op}(s,t) \, ,
  \label{opdef}
  \eea
  that contains poles in the $s$ and $t$ channels.  The factor $P_4$ is a dimension-four kinematic prefactor that is determined by maximal Yang--Mills supersymmetry and contains the information about which particular   states are being scattered.  For example,  in the case of the four-gluon amplitude this prefactor is given by $P_4=F^4$, where $F$ is the linearised field  strength.\footnote{The manifestly supersymmetric amplitude from which the component  expression \eqref{opdef} arises can be  expressed as $\delta^8(Q_4) \, A_{op}(s,t) $, where $Q_4$ is the supercharge for  the four scattering states.} 
 
 The amplitude \eqref{opdef} contains the stringy corrections to the field theory tree-level amplitude, which is $P_4/(st)$. In order to obtain an expression that is well-defined at $t=0$, we  will  again consider the amplitude  after subtraction of the super Yang--Mills tree-level term,  
\begin{align}
\tilde A_{op}(s,t)  :=  A_{op}(s,t) + \frac{1}{st} &= -\frac{1}{st} \,\left( \frac{\Gamma(1- s)\Gamma(1- t) } {\Gamma(1-  (s+t)) }-1\right)  \nn\\
&=- {1 \over s\, t}  \left( \exp\left[ \sum_{k=2}^{\infty} {\zeta(k) \over k} ( s^k + t^k -(s+t)^k) \right] -1 \right)\,.
\label{subtract}
\end{align}
From the expression in the second line  it is obvious that $\tilde A_{op}(s,t) $ has a low energy expansion in powers of $s$ and $t$ with coefficients that are rational polynomials in Riemann zeta values.
The amplitude $\tilde A_{op}(s,t)$ can also be written as a sum of $s$-channel poles by using the integral representation for the Euler beta function,
\begin{align}
\tilde A_{op}(s,t) 
&=   \frac{1}{t} \int_0^1 dx \, x^{-1-s}\, (1-x)^{-t} + \frac{1}{st} \nn\\
&=  \frac{1}{t} \int_0^1 \frac{dx}{x} \sum_{m=1}^\infty  x^{-s+m} \, (-1)^m \, {-t\choose m} \nn\\
&=  \sum_{m=1}^\infty \frac{1}{m-s} \,\frac{\Gamma(m+t) }{\Gamma(1+t) \Gamma(1+m) } =  \sum _{m=1}^\infty \sum_{\ell_m=0}^{m-1}\frac{p_m^{\ell_m} \, G_{\ell_m}^{\frac{D-3}{2}}   \left(1+\frac{2t}{m}\right)) }{m-s}\,.
 \label{openamp}
\end{align}
 Writing the amplitude in this manner exhibits the infinite set of poles at positive integer values of $s$, but obscures the $s-t$ symmetry of the amplitude and, in particular, obscures the presence of an infinite set of poles at positive integer values of $t$.  
 The last equality of \eqref{openamp} expresses the amplitude  in the form  of a partial wave sum over Gegenbauer polynomials of the same form as  $A^{(s)}_4(s,t)$ in \eqref{genamp} (where $m_a^2$ takes integer values, $m$, and the angular momentum of states at mass $m$ takes the values, $0\leq \ell_m\le m+1$).    The identity in the last line only implies $p_m^{\ell_m}\ge 0$ when $D\le 10$, which is consistent with the no-ghost theorem.  A more complete derivation of the amplitude \eqref{subtract} requires the condition $D=10$ of the critical superstring.

The asymptotic behaviour of $\tilde A_{op}(s,t)$ as $|s|\to \infty$ with $\epsilon <\arg(s) < 2\pi-\epsilon$ , and $t\leq 0$, can be obtained  by using Stirling's approximation, giving
\bea
\tilde A_{op}(s,t) &\underset{s\to \infty}{\sim} &   {1\over s \, t}  - (-s)^{t-1} \left(\Gamma(-t) +O(s^{-1})\right)\, ,
\label{asymbeta}
\eea
so the last term in \eqref{cauchy}  (the large-$s$ contour integral) can be dropped.

\subsection{Low-energy expansion of the four-particle open superstring tree  amplitude}  

The expansion of the expression \eqref{openamp} in powers of $s$ and $t$ is straightforward and has the form
\bea
\tilde A_{op}(s,t) =\sum^{\infty}_{p,q=0} g^{op}_{p,q} \, s^{p} t^q\, .
\label{firstop}
\eea  
The coefficients  in this expansion, $g^{op}_{p,q}$, are rational polynomials in Riemann zeta values with weights $w=p+q+2$.   The terms up to order $t^3$ and order $s^5$ are shown in \eqref{openexp}.

\bigskip
{\it The $(p,0)$ terms.}
\smallskip

The leading power of $t$ is picked out by considering the forward limit, $t=0$, in which case the amplitude reduces to the  simple form
 \begin{align}
  \tilde A_{op}(s,0) &=
   \sum_{n=0}^\infty \zeta(n+2) \,  s^n \nn\\
 &=  -\frac{1}{s}(\gamma+ \psi(1-s) )\nn\\
   &= \sum_{n=1}^\infty \frac{1}{n(n-s)} = \frac{1}{s} \sum_{n=1}^\infty  \left(\frac{1}{n-s}-\frac{1}{n}\right)\,,
 \label{tzero}
\end{align}
 where  the digamma function is defined by $\psi(z) =\Gamma'(z)/\Gamma(z)$ and 
the Euler--Mascharoni constant is defined by $\gamma = - \psi(1) = - \Gamma'(1)$.  
It follows that in  this case  the coefficients, $g^{op}_{p,0}$ are simply given by
 \bea
 g^{op}_{p,0}= \zeta(p+2) \, .
 \label{zerot}
 \eea
 
 We also see from \eqref{tzero}  that the $t=0$ contribution  may be expressed as an infinite sum of poles with positive residues in accord with unitarity. 
 It follows from our previous discussion that the determinants of the $n\times n$ Hankel matrices with $\zeta(i+j)$ entries (where $i$ is even or odd) are all positive, as are all the minors of these matrices.  

 Such determinants of Hankel matrices with $\zeta$-value entries have been considered in the mathematics literature \cite{Monien:2009,Haynes:2015}. These references considered the behaviour of the determinants of the $2n\times 2n$  Hankel matrices
 \bea
\det   H^{(0)}_{n}[\zeta]_{ij} =\det( \zeta(i+j) )\,, \qquad i,j=1,2,\dots,n \, ,
 \label{Hank1}
 \eea
 and 
 \bea
\det H^{(1)}_{n}[\zeta]_{ij} = \det(\zeta(i+j+1)) \,,  \qquad i,j=1,2,\dots, n\,  ,
 \label{Hank2}
 \eea
 which were both argued to be positive.  This follows from the fact that $\det H_n^{(1)}[\zeta]_{ij}$ is a principal minor of  $H_{n+1}^{(0)}[\zeta]_{ij}$.  In fact, as we commented previously, it is a property of the Stieltjes moment coefficients that if two such Hankel matrices are known to have positive determinants,   all other minors are positive  \cite{independent-Hankel}.

 It is easy to see that $\det H^{(i)}_{n}[\zeta]$ ($i=0,1$)  approaches zero rapidly as  $n$  increases. More explicitly, it was reported in  \cite{Monien:2009,Haynes:2015} that the asymptotic values of these determinants at large $n$ are given by  the expressions\footnote{The constants $d^{(0)}$ was denoted $A^{(0)}$ in \cite{Monien:2009}.}
\bea
\det H^{(0)}_{n}[\zeta] = d^{(0)} \left(\frac{2n+1}{e\sqrt e}\right)^{-(n+1/2)^2}\left(1+\frac{1}{24} \frac{1}{(2n+1)^2}+\dots\right)\, ,
\label{dethank0}
\eea 
and
\bea
\det H^{(1)}_{n-1} [\zeta] = \frac{e^{9/8}}{\sqrt 6}\, d^{(0)} \, \left(\frac{2n}{e\sqrt e}\right)^{-n^2+3/4}\left(1-\frac{17}{240} \frac{1}{(2n)^2}+\dots\right)\, ,
\label{dethank1}
\eea 
It is easy to check these expressions with help from Mathematica, although we find the numerical constant $d^{(0)}=0.66367$ rather than the value attributed to Zagier in \cite{Monien:2009}, which is $d^{(0)}= 0.35147$.
 
\bigskip

{\it The $(p,1)$ terms.}
\smallskip

It is also easy to see that the $q=1$ terms (terms of order $t^1$) are given by 
\bea
\partial_t \tilde{A}^{(1)}_{op}(s,t)|_{t=0} = \sum_{p=0}^{\infty} g^{op}_{p,1} \, s^{p} \, ,
\eea
with 
\bea
g^{op}_{p,1} =  {p+2 \over 2} \zeta(p+3) -{1\over 2} \sum_{i=1}^{p} \zeta(i+1) \zeta(p+2-i) \, .
\eea  
In this case one may construct Hankel matrices $H_n^{op\,(0)}[{\bf g_1^{op}}]$ of the form \eqref{hankdef} with entries $\{ g^{op}_{0,1},\dots,g^{op}_{n,1}\}$ in the first row.  The determinants of such matrices and all their minors again satisfy positivity conditions.    Since $g_{p,1}^{op}$ is quadratic in zeta values such bounds now imply more complicated bounds on rational  polynomials of zeta values.

\bigskip
{\it All $(p,q)$ terms.}
\smallskip

It is tedious to extract the complete set of coefficients for $q>1$ simply by expanding the expression in \eqref{subtract}.  However, it was  shown recently \cite{Zagier:2019eus} (see also \cite{Hoffman:1997})   that the coefficient $g^{op}_{p,q}$ is given by the special multiple zeta values\footnote{A general multiple zeta value of depth $r$ and weight $w=\sum_{i=1}^r k_i$ is defined by $\zeta(k_1,k_2,,\dots, k_r)=\sum_{0<n_1<\dots <n_r} n_1^{-k_1}\dots n_r^{-k_r}$ }  (which can also be expressed as special kinds of Mordell--Tornheim sums)
\bea
g^{op}_{p,q}= \zeta({\underbrace{1, \cdots, 1}_{q}}, p+2)\, .
\label{gmzv}
\eea
These expressions can be reduced to rational polynomials in single zeta values of total weight $p+q+2$ by 
comparing the coefficient of $s^pt^q$ in the low energy expansion of the open-string amplitude \eqref{subtract}  with \eqref{gmzv}.  In this manner we have been able to determine the following expressions for  the coefficients with $q=0,\dots,  3$ for all $p\ge0$,
\begin{align} \label{single-zeta}
& q=0: \quad \zeta(p+2) \, , \\
& q=1: \quad \zeta(1, p+2) =  {p+2 \over 2} \zeta(p+3) -{1\over 2} \sum_{i=1}^{p} \zeta(i+1) \zeta(p+2-i) \, , \cr
& q=2: \quad  \zeta(1,1, p+2) =   {(p+2)(p+3) \over 3!} \zeta(p+4) -  \sum_{i=1}^{p} {i+1\over 2!} \zeta(i+2) \zeta(p+2-i) \cr
&~~~~~~~~~ ~~~~~~~~~~~ ~~~~~~~~~ + {1\over 3!} \sum_{\underset{ i+j\leq p} {i,j=1}}^{p} \zeta(i+1)\zeta(j+1)\zeta(p+2-i-j)\, , \cr
& q=3: \quad  \zeta(1,1,1,  p+2) = {(p+2)(p+3)(p+4) \over 4!} \zeta(p+5) - \sum^p_{i=1} {(i+1)(i+2) \over 3!} \zeta(i+3)\zeta(p+2-i)
\cr 
& ~~~~~~~~~ - \frac{1}{2!} \sum^p_{i=1} {i+1 \over 2! }{p+2-i \over 2! } \zeta(i+2)\zeta(p+3-i) 
+\frac{1}{2!}\sum^p_{\underset{ i+j\leq p} {i,j=1}} {j+1 \over 2!} \zeta(i+1)\zeta(j+2) \zeta(p+2-i-j)  \cr
& ~~~~~~~~~ - {1\over 4!}  \sum^p_{\underset{ i+j+k\leq p} {i,j, k=1}} \zeta(i+1)\zeta(j+1)\zeta(k+1) \zeta(p+2-i-j-k)  \, .  \nonumber
\end{align} 

The class of $n\times n$ Hankel matrices that is generated from the low-energy expansion of the open-string four-particle amplitudes at a given order $t^q$  is given by
\bea
H^{op}_{n} [{\bf \zeta_q}]_{ij}  =  \zeta({\underbrace{1, \cdots, 1}_{q}}, i+j)    \,,  \qquad i,j=1,2,\dots, n \, ,
\eea
where the matrices with  $n=1, 2, 3, \cdots$. are sub-matrices of the infinite-dimensional matrix $H^{op}_{\infty} [\zeta]_{ij}$.
The notation $H^{op}_{n} [{\bf \zeta_q}]$  denotes  the Hankel matrix with the first row defined by the sequence $\{ \zeta({\underbrace{1, \cdots, 1}_{q}}, 2),\dots,  \zeta({\underbrace{1, \cdots, 1}_{q}}, n+1)\}$.
For example, the Hankel matrices with $n=3$ and general $q$ have the form 
 \bea
H^{op}_{3} [\bf{\zeta_q}]= \begin {pmatrix}
\zeta({\underbrace{1, \cdots, 1}_{q}}, 2) &  \zeta({\underbrace{1, \cdots, 1}_{q}}, 3) & \zeta({\underbrace{1, \cdots, 1}_{q}}, 4)   \\
\zeta({\underbrace{1, \cdots, 1}_{q}}, 3) &  \zeta({\underbrace{1, \cdots, 1}_{q}}, 4) & \zeta({\underbrace{1, \cdots, 1}_{q}}, 5)   \\
\zeta({\underbrace{1, \cdots, 1}_{q}}, 4) &  \zeta({\underbrace{1, \cdots, 1}_{q}}, 5) & \zeta({\underbrace{1, \cdots, 1}_{q}}, 6)     
  \cr
 \end{pmatrix} \, .
 \label{hankdef2}
 \eea
The unitarity constraints again imply that 
\bea
\det H^{op}_{n} [{\bf \zeta_q}]  > 0 \, ,\eea
for all $n\ge 1$ and $q\ge 0$, as well as similar positivity constraints on all the minors.
The simplest example of  many such constraints comes from the positivity of the upper left  $2\times 2$ minor
\bea
\zeta({\underbrace{1, \cdots, 1}_{q}}, 2) \,  \zeta({\underbrace{1, \cdots, 1}_{q}}, 4)  - ( \zeta({\underbrace{1, \cdots, 1}_{q}}, 3) )^2
>0\,.
\label{twotwo}
\eea
This is just one of an infinite number of positivity  constraints that can be reduced to inequalities on polynomials of positive zeta values.  It is straightforward to check these numerically and to check that $\det H^{op}_{n} [{\bf \zeta_q}] $ decreases rapidly to zero as $n$ grows.  However, we have not obtained expressions analogous to \eqref{dethank0} and \eqref{dethank1}, which would give the asymptotic dependence of $\det H^{op}_{n} [{\bf \zeta_q}] $  on $n$.

 \section{Four-particle closed superstring tree amplitudes}
 \label{sec:closed}
 
The four-particle  closed-string tree amplitude has the form 
\bea
\cA_{cl}(s,t) =  P_8\,   A_{cl}(s,t)\, ,
\label{closedef}
\eea where $P_8$ is a dimension-eight kinematic factor that is determined by supersymmetry,  such as $R^4$ in the case of the four-graviton amplitude (where $R$ is the linearised Riemann curvature).\footnote{The prefactor $P_8$ is the component expression corresponding to the manifestly supersymmetric prefactor  $\delta^{16}(Q_4)$ that enters the superamplitude that  describes the scattering of any four massless states in the  gravity supermultiplet with maximal supersymmetry.}  
The factor $A_{cl}(s,t)$ is given by 
 \bea
 A_{cl}(s,t) &=&- \frac{1}{st(s+t)}\frac{\Gamma(1- s)\Gamma(1- t) \Gamma(1+ (s+t))} {\Gamma(1+ s)\Gamma(1+ t) \Gamma(1-  (s+t)) }  \nn\\
 &=& - {1 \over s t (s+t)}  \exp\left[ \sum_{k=2}^{\infty} {2\zeta(2k+1) \over 2k+1} ( s^{2k+1} + t^{2k+1} -(s+t)^{2k+1}) \right]
 \,,
 \label{closed}
 \eea
 where the second expression is useful for exhibiting the low-energy expansion. It follows that the first term in this expansion is  $P_8/(st u)$,  which contains the tree-level supergravity  four-particle amplitudes.  Once again we will avoid the $t=0$ singularity by subtracting the classical term by defining
 \bea
\tilde A_{cl}(s,t) :=  A_{cl}(s,t) +\frac{1}{s t  (s+t)}\,.
 \label{closedpole}
 \eea
  In this case the amplitude not only has poles on the positive real $s$ axis but also on the positive $u$ (i.e., negative $s$) axis and is a special case of the general  structure in \eqref{fullamp}.
It is easy to see that for $t\le 0$ this expression has the asymptotic behaviour $\tilde A_{cl}(s,t) \underset{s \to -\infty  } {\sim}(-s)^{-2}$ at fixed $t$,  which means that the boundary term in \eqref{cauchy} can be dropped. 

We will now express $  \tilde A_{cl}(s,t) $ as a sum of $s$-channel and $u$-channel poles in the form 
 \bea
  \tilde A_{cl}(s,t)  =    \tilde  A^{(s)}_{cl}(s,t) +   \tilde  A^{(u)}_{cl}(u,t) \,,
 \label{regions}
 \eea
making use of the integral representation (which was used in the original paper by Shapiro \cite{Shapiro:1970gy})
\bea
\int_\CC d^2w |w|^{-2-2s}|1-w|^{-2t}& =& -\frac{\pi t}{s (s+t)} \frac{\Gamma(1- s)\Gamma(1- t) \Gamma(1+ (s+t))} {\Gamma(1+ s)\Gamma(1+ t) \Gamma(1-  (s+t)) }\nn\\
&=&\pi t^2    A_{cl}(s,t) \,.
  \label{virasoro}
  \eea
Dividing the integration domain into the regions (1) $|w|\le 1$ and (2)  $|w|\ge 1$ and using the fact that 
  \bea
  \int_{|w|\ge 1}  d^2w |w|^{-2-2s}|1-w|^{-2t} = \int_{|w|\le 1} d^2w |w|^{-2-2u}|1-w|^{-2t}\,,
  \label{wrels}
\eea 
it follows that  region (2) is equivalent to region (1) with $s\to u$.  

We now isolate the contributions from the $s$-channel and $u$-channel poles in 
  \begin{align}
   \tilde A^{(s)}_{cl}(s,t) & := \frac{1}{\pi t^2} \int_0^1 \frac{ dr}{r} \int _0^{2\pi}  d\theta\,  r^{-2s} (1-r e^{i\theta})^{-t}(1-r e^{-i\theta})^{-t} + \frac{1}{t^2 s} \nn\\
   &=  \frac{2}{t^2} \int_0^1 \frac{dr}{r} \sum_{m=1}^\infty  r^{-2s+2m}  \, {-t\choose m}^2 \nn\\
   &= \sum_{m=1}^\infty \frac{1}{m-s} \left(\frac{\Gamma(m+t) }{\Gamma(1+t) \Gamma(1+m) }\right)^2\,,
  \label{rless1}
  \end{align}
  and
  \begin{align}
   \tilde A^{(u)}_{cl}(u,t)  &:= \frac{1}{\pi t^2} \int_0^1 \frac{ dr}{r} \int _0^{2\pi}  d\theta\,  r^{2s+2t} (1-r e^{i\theta})^{-t}(1-r e^{-i\theta})^{-t} + \frac{1}{t^2u} \nn\\
   &=   \sum_{m=1}^\infty \frac{1}{s+t+m} \left(\frac{\Gamma(m+t) }{\Gamma(1+t) \Gamma(1+m) }\right)^2 \,,
   \label{rless2}
  \end{align}
 where  we have used the fact that 
  \bea
  \frac{1}{t s(s+t) }  = \frac{1}{t^2 s} + \frac{1}{t^2 u} \, .
  \label{poledefs}
  \eea 
Comparing the expressions \eqref{rless1}  and \eqref{rless2} with  \eqref{openamp}, it is apparent that the closed string tree amplitude can be obtained from the open string amplitude by squaring  the residue of each pole. This is closely related to the Kawai, Lewellen and Tye  (KLT) relation \cite{Kawai:1985xq}
and is the string generalization of  the Bern, Carrasco and Johansson (BCJ) double copy~\cite{Bern:2008qj} relation between gauge theory and gravity.

The positivity conditions on the $s$-channel contribution $\partial_t^q\tilde A^{(s)}_{cl}(s,t)\big|_{t=0}$, follow from the fact that the residues of the massive poles  have positive coefficients when expanded in terms of Gegenbauer polynomials.   We now need to check  the asymptotic forms  of $A^{(s)}_{cl} (s,t)$ and $A^{(u)}_{cl}(u,t)$ when $t$ is fixed  and $|s|\to \infty$.     
  We have (setting $r=e^{-y}$),  
  \begin{align}
 A^{(s)}_{cl}(s,t) &=  \frac{1}{\pi t^2} \int_0^\infty dy\int _0^{2\pi}  d\theta\,  e^{2sy} (1-e^{-y} e^{i\theta})^{-t}(1-e^{-y} e^{-i\theta})^{-t}  \nn\\
  &=  \frac{1}{\pi t^2} \int_0^\infty dy\int _0^{2\pi}  d\theta\,  e^{2sy} (1+e^{-2y} - 2 e^{-y} \cos \theta )^{-t}  \nn\\
  &\underset{s\to \infty} {\to}   \frac{2^{-t}}{\pi t^2 s} \int_0^\infty d\hat y\int _0^{2\pi}  d\theta\,  e^{-2\hat y} ((1  -  \cos \theta) (1+\frac{\hat y}{s}) )^{-t} \nn\\
    &=  \frac{ 2^{-2t} \,\Gamma \left(\frac{1}{2}-t\right)}{\pi^{\frac{1}{2}}\, s\, t^2\,  \Gamma (1-t)} + {\cal O}(s^{-2}) \, ,
  \label{asmone}
  \end{align}
where we have rescaled   $y\to \hat y/(-s)$ to account for the limit $s\to -\infty$.  We have then expanded the factor of $e^{-y} =e^{\hat y/s} \sim 1+\hat y/s+O(s^{-2})$ in the two brackets.
We therefore deduce that 
  \begin{align}
 \tilde A^{(s)}_{cl}(s,t) = A^{(s)}_{cl}(s,t) -\frac{1}{s t^2}  \underset{s\to \infty} {\to} & \frac{1}{s t^2}\,  \left(\frac{ 2^{-2t} \,\Gamma \left(\frac{1}{2}-t\right)}{\pi^{\frac{1}{2}}\,  \Gamma (1-t)} -1\right)+O(s^{-2})        \,.
\label{atildasym}
  \end{align}
More generally, the large-$|s|$ expansion of the amplitude has the form $ \tilde A^{(s)}_{cl}(s,t)   \underset{s\to \infty} {\to}  \sum_{q=0} c_q t^q\,s^{-1}$  with $q\ge 0$, where $c_q$ is constant.      Therefore, the contour integral at $|s|\to \infty$ in \eqref{cauchy} vanishes.
The $1/s$ behaviour in (\ref{asmone}) cancels with the leading term in  $ A^{(u)}_{cl}(u,t)$ in the complete amplitude.
The full amplitude is Regge behaved and  behaves as  $(-s)^{t-2}$ as $|s|\to \infty$.  
 
 \subsection{The closed-string low-energy expansion coefficients}
 
 In order to discuss the low-energy expansion of the closed-string tree amplitude  we first note that the terms of lowest-order in $t$, i.e. the expansion of the $t=0$ amplitude, have the form 
  \begin{align}
  \tilde A_{cl}(s,0) &= \tilde A^{(s)}_{cl}(s,0) +  \tilde A^{(u)}_{cl}(s,0) =    2 \sum_{n=0}^\infty \zeta(2n+3) \, s^{2n} \cr
&=-\frac{1}{s} \left(2 \gamma +\psi(1-s)+\psi(1+s)\right)  \nn\\
&= \frac{1}{s^2} \sum_{k=1}^\infty\left( \frac{ 1}{k-s} + \frac{1}{k+s}- \frac{2}{k} \right)    \,,
 \label{zerocl}
 \end{align}
 which is  simply a constant plus  the sum of $s$-channel and $u$-channel poles. In this case the coefficient of $s^{2n}$ is $g_{2n,0}= 2\, \zeta(2n+3)$.  The relevant Hankel matrices
   can be viewed as sub-matrices  of the open-string Hankel matrixes, with the even zeta values set to zero and with $\zeta(2n+1)\to 2 \zeta(2n+1)$.  This is the result of the single-valued projection \cite{Brown}, which also reflects the KLT  relation between open and closed string tree amplitudes (see e.g. \cite{Brown:2018omk, Schlotterer:2018zce, Vanhove:2018elu} for recent applications to superstring amplitudes).  The positivity  of the determinant of the Hankel matrices   formed from these coefficients provides no further constraints on products of zeta values beyond those deduced from the open-string case.

\subsection{Low-Energy Expansion of $\tilde A^{(s)}_{cl}(s,t)$} 

Unlike the colour-ordered open string amplitudes, the coefficients of low-energy expansion $\tilde A_{cl}(s,t)$ are generally not positive definite  due to the $u$-channel contribution,  Consequently,  they do not reside on a moment curve, apart from the  $t^0$ term discussed earlier. To deal with this issue, we will consider the unitarity constraints on the $s$-channel contribution, $\tilde A^{(s)}_{cl}(s,t)$, in (\ref{rless1}). As we will show below, the coefficients of  the low-energy  expansion of $\tilde A^{(s)}_{cl}(s,t)$ are not only positive, but also satisfy the Hankel matrix constraints, just as in the case of open superstring amplitudes. Interestingly, even though the low-energy expansion of $\tilde A_{cl}(s,t)$ only contains powers of single odd  zeta values, individually the low-energy coefficients of $\tilde A^{(s)}_{cl}(s,t)$ and $\tilde A^{(u)}_{cl}(u,t)$  include irreducible MZVs as well as even zeta values. Therefore, the unitary of $\tilde A^{(s)}_{cl}(s,t)$ leads to Hankel matrix constraints on irreducible MZVs.

 \subsubsection*{The low energy expansion coefficients.}

The expansion of $\tilde A^{(s)}_{cl}(s,t)$ and $ \tilde A^{(t)}_{cl}(u,t)$ can be obtained to any given order in the low energy expansion by explicit expansion of \eqref{rless1} and \eqref{rless2}.   Motivated by the expressions in \cite{Zagier:2019eus} we  obtain this expansion in the form
  \begin{align}
 \tilde A^{(s)}_{cl}(s,t)   &=  \sum_{n=1}^\infty \frac{1}{n-s} \left(\frac{\Gamma(n+t) }{\Gamma(1+t) \Gamma(1+n) }\right)^2 =  \sum_{p=0}^\infty s^p \sum_{n=1}^\infty \frac{1}{n^{p+1}} \left(\frac{\Gamma(n+t) }{\Gamma(1+t) \Gamma(1+n) }\right)^2 \nn\\
    &= \sum_{p=0}^\infty s^p \sum_{q=0}^\infty  Z(p+3, q)\, t^q \,.
 \label{zclosed}
 \end{align}
 
The quantity $Z(p+3,q)$ is defined by the generating function 
\bea
\sum_{q=0}^\infty Z(p+3,q) \,  t^q = \sum_{n=1}^\infty\frac{1}{n^{p+1}} \left(\frac{\Gamma(n+t) }{\Gamma(1+t) \Gamma(1+n) }\right)^2 = \sum_{n=1}^\infty \frac{1}{n^{p+3}}\prod_{0<m<n}\left (1+\frac{t}{m} \right)^2\, ,
\eea
and can be expressed as a linear sum of elements of a particular class of  MZVs of weight $q+r$, 
\bea
Z(r,q):=\sum_{\underset{q_1+\dots + q_j=q} {q\in \{1,2\}^j, \, j\ge 0}} 2^{\#\{i:q_i=1\}} \,\zeta({\bf q},r)\,.
\label{zdef}
\eea
In this expression the components of the $j$-component vector ${\bf q} = (q_1,q_2,\dots,q_j)$ are summed over values $q_i=1,2$, subject to the condition $\sum_{i=1}^j q_i =q$. Furthermore, $\#\{i:q_i=1\}$ in the coefficient $2^{\#\{i:q_i=1\}}$ denotes the numbers of components with $q_i=1$.
It follows from \eqref{zclosed} that  the closed-string coefficients are simply given by  
 \bea
 g^{cl}_{p,q} =  Z(p+3, q)\, . 
 \label{coeffsclosed}
 \eea

Interestingly, as stressed in \cite{{Zagier:2019eus}},  the quantities $Z(r,q)$ not only arise in the expansion of the tree-level closed-string amplitude, but also in the evaluation of the low-energy expansion of the genus-one four-graviton amplitude \cite{Green:2008uj}.  In that context certain multiple sums arise in considering the coefficients in the Laurent polynomial of the  large-$\Im \tau$ expansion  of the two-point functions, $D_\ell(\tau)$, on a genus-one surface of   complex structure $\tau$.  These multiple sums have the form 
 \bea
S(m,n)\equiv \sum_{k_1,\dots,k_m\ne0}  \frac{\delta(\sum_{1\le i\le m}k_i)}{|k_1\cdots k_m|(|k_1|+\cdots+|k_m|)^n} \, .
\label{zdefs}
\eea
In appendix A.3.2 of \cite{Green:2008uj}  authored by Don Zagier it was proved that 
\bea
S(q+2,r-2)= (q+2)! \, 2^{2-r}\, Z(r,q) \, ,
\label{zdefs2}
\eea
with $Z(r,q)$ given by \eqref{zdef}.  The interesting  fact that $Z(r,q)$  arises in the low-energy expansion of the genus-one amplitude as well as the expansion of the  tree amplitude was emphasised in \cite{Zagier:2019eus}.

The expressions for the coefficients  of $t^0$ and $t^1$ are  the following combinations of polynomials in zeta values
\begin{align}
Z(r,0) &= \zeta(r)\,,\nn\\
Z(r,1)&= 2\, \zeta(1,r)= {r}\, \zeta(r+1) -  \sum_{i=1}^{r-2} \zeta(i+1)\,\zeta(r-i)\, . 
\label{zexs}
\end{align}
We see that the coefficients of  the low-energy expansion of $\tilde A^{(s)}_{cl}(s,t)$  bear a very close resemblance to the coefficients in the expansion of the  open-string amplitude in  \eqref{single-zeta}.  The  $q=0$ (i.e. $t^0$) terms are identical whereas the $q=1$ (i.e. $t^1$) terms have an additional factor of $2$. This fact can be understood from the double copy structure of \eqref{rless1} so  that the residue of each pole of closed-string amplitude is the square of that of the open string amplitude.

When $q=2$, we have
\bea
Z(r,2)=\zeta(2,r)+4\, \zeta(1,1,r) \,,
\label{multiz}
\eea 
where $\zeta(1,1,r)$ can always be reduced to zeta values  as given in  \eqref{single-zeta}, and $\zeta(2,r)$ can also be reduced to a polynomial in zeta values when $r<6$.  However, when $r= 6$ the MZV $\zeta(2,6)$ is irreducible.  Indeed, $Z(r,q)$ generally contains irreducible MZVs when $q>1$ and $r+q\ge 8$. The  $\tilde A^{(u)}_{cl}(u,t)$ part of the complete closed-string amplitude has the same structure as above, but with $s$ replaced by $u=-s-t$. All the  even zeta values as well as all the irreducible MZVs cancel in the sum of  $\tilde A^{(s)}_{cl}(s,t)$ and $\tilde A^{(u)}_{cl}(u,t)$ so the full four-particle closed-string tree amplitude can be expressed in terms of odd zeta values only. 

It follows from the earlier discussion that the $n \times n$ Hankel matrix associated with the coefficients of the low-energy expansion of $\tilde A^{(s)}_{cl}(s,t)$ has elements given by 
 \bea
 {H^{cl\, (s)}_{n}} [{\bf   Z_q} ]_{ij}  =  Z( i+j +1,q)   \,,  \qquad i,j=1,2,\dots, n \, ,
 \label{hankclosed}
 \eea    
 An example of a  $3\times 3$ Hankel matrix for any $q$ is given by  
  \bea
H^{cl\, (s)}_{3} [\bf{Z_q}]= \begin {pmatrix}
  Z(3, q) &  Z(4, q) &  Z(5, q)   \\
  Z(4, q)  & Z(5, q) & Z(6, q) \\
  Z(5, q) &  Z(6, q) &  Z(7, q)  
 \end{pmatrix} \, .
 \label{hankdef3}
 \eea
 Much as before, the positivity conditions lead to conditions of the form
 \bea
 \det\, H^{cl\, (s)}_{n}[ {\bf  Z_q }] > 0\, ,
 \label{closedpos}
 \eea
 as well as a host of such inequalities expressing the positivity of any  minor of ${H^{cl\, (s)}_{n,q}}[Z]_{ij}$ for all  $n\ge3$ 
 . As discussed earlier, when $q\geq 2$, $Z( i+j +1,q)$ generally contains irreducible MZVs. For  example,  in the $q=2$ case \eqref{hankdef3} becomes
\bea
\!\!\!\!\!\!\!\!\! H^{cl\, (s)}_{3} [{\bf  Z_2}] = \begin {pmatrix}
\frac{5\zeta(5)}{2} {-}  \zeta(2) \zeta(3) &  {53  \zeta(6) \over 12} {-} 3 \zeta(3)^2 &  9\zeta(7) {-} 3\zeta(2)\zeta(5) {-} 3\zeta(3)\zeta(4)  \\
\ldots & \ldots & \zeta(2,6) {+} {61 \zeta(8) \over 6}   {+} 2 \zeta(2) \zeta(3)^2 {-} 12 \zeta(3) \zeta(5) \\
\ldots&  \ldots &  {113 \zeta(9)\over 6} {-} 5\zeta(2)\zeta(7) {-} 5\zeta(3)\zeta(6) {-} 5\zeta(4)\zeta(5) {+} {2\zeta(3)^3\over 3}  
  \cr
 \end{pmatrix}\nn\\
 \eea
 where  the ellipsis represent entries that are  identified by the fact that the matrix is symmetric.  In this case the positivity conditions \eqref{closedpos} lead to conditions on polynomials that contain the irreducible  MZV $\zeta(2,6)$.

\section{ Discussion} 
\label{sec:discussion}

As discussed by Arkani-Hamed, Huang and Huang \cite{EFThedron, ArkaniHamed:2018,Huang:2018}, general considerations of unitarity and asymptotic behaviour of four-particle scattering amplitudes lead to very interesting geometric constraints on low-energy physics. As a consequence the coefficients in the low-energy expansion must reside inside a cyclic polytope, which is determined by the Gegenbauer polynomials.  This  leads to a large number of positivity constraints on polynomials of the low energy coefficients that are encoded in the positivity of Hankel matrices and their minors.

 In this paper we have explored these positivity constraints on the coefficients of the  low-energy expansions of tree-level amplitudes in open and in closed superstring theories.   Despite the fact that considerations of the four-particle amplitude with massless external states  can only probe a limited  amount of information  this nevertheless leads to a host of interesting inequalities involving coefficients in the low energy expansion, which are rational polynomials of multiple zeta values.    These constraints follow from positivity of the determinants of the Hankel matrix (and any of its minors) formed from  these  coefficients. The simplest version of these inequalities reproduces the known results in math literature \cite{Monien:2009, Haynes:2015}. Our consideration from unitarity of superstring amplitudes not only provides a physical interpretation of the known inequalities among single zeta values,  but also leads to a host of new relations among single zeta values as well as more general MZVs. These inequalities are the necessary conditions for superstring amplitudes being unitary, it would be of interest to prove these inequalities by other means.

 Another aspect of the positivity properties of the amplitude that was stressed in \cite{EFThedron, ArkaniHamed:2018,Huang:2018} involves  reorganising the low-energy expansion so that it takes the form $\sum_{\Delta,q} \tilde g_{\Delta,q} s^{\Delta-q}\, t^q$, where $\Delta=p+q$.  This leads to positivity conditions on the vector of coefficients, 
 ${\bf g_{\Delta}} =\{g_{\Delta,0},\dots ,g_{\Delta,n}\}$ that imply that this vector (for arbitrary $n$)  must reside inside the cyclic polytope generated by expanding the Gegenbauer polynomials in  powers of $t$~\cite{EFThedron, ArkaniHamed:2018,Huang:2018}.  
It might be interesting to study the implications of these constraints on the MZV coefficients of open and closed string theories.

It is of note that in the case of the closed-string amplitude, the positivity constraints are constraints on rational  polynomials of irreducible MZVs.  These follow from the introduction of the quantities $Z(r,q)$ (introduced in  \cite{Zagier:2019eus} and defined in \eqref{zdef}) that are combinations of MZVs that arise as intermediate coefficients in the low energy expansion of the closed-string four-particle amplitude.  Although the irreducible MZVs cancel in the expansion of the full four-particle amplitude, they contribute to the portion  of the amplitude that has $s$-channel poles and satisfy the conditions contained in \eqref{closedpos}.

The inequalities satisfied by polynomials in  MZVs implied by  unitarity of superstring tree  amplitudes generalise  results in the mathematics literature \cite{Monien:2009, Haynes:2015}   on determinants of Hankel matrices of single zeta values. 
The  determinants of these  $n\times n$ Hankel matrices approach zero very rapidly as  a function of $n$. This is easily verified by direct numerical  estimation and in the cases with single zeta values the explicit expressions for the  asymptotic behaviour  are known \cite{Monien:2009, Haynes:2015} (and are quoted in  \eqref{dethank0} and \eqref{dethank1}).
It would be of interest  to determine analogous expressions for the $n$-dependence of the asymptotic behaviour of the determinants of Hankel matrices of MZVs that arise in this paper.  It would also  be of obvious interest to develop an interpretation of the asymptotic behaviour of such matrices in terms of  asymptotic properties  of superstring scattering amplitudes.

We know   that in order to resolve the full content of the no-ghost theorem  \cite{Brower:1972wj,Goddard:1972iy,Thorn:1983cz} it is necessary to consider massless $N$-point amplitudes for all values of $N$.  This should be possible, given the explicit expressions for such amplitudes in both open and closed superstring theories \cite{Schlotterer:2012ny,Mafra:2011nv}.  The study of  higher-point massless amplitudes or  four-point amplitudes with more general massive external states should lead to more general unitarity conditions on the MZVs. 
Furthermore, the generalisation to amplitudes of higher genus raises interesting  new issues relating to the presence of massless threshold singularities that arise in the low energy expansion (such as those discussed in the genus-one case in  
\cite{Green:2008uj,DHoker:2019blr}).

\section{Acknowledgements}

We are grateful to  Oliver Schlotterer   for useful conversations and to Nima Arkani-Hamed for comments on the draft of this paper. CW would like to thank Yu-tin Huang for collaboration and discussions on related topics. CW is supported by a Royal Society University Research Fellowship No. UF160350.   MBG has been partially supported by STFC consolidated grant ST/L000385/1 and by a Leverhulme Emeritus Fellowship.    This work was  finalised  during the workshop ``Scattering Amplitudes and the Conformal Bootstrap" at the Aspen Center for Physics, which is supported by National Science Foundation grant PHY-1607611, and  was partially supported by a grant from the Simons Foundation.

\appendix
 
 \section{Four-particle superstring tree  amplitudes at higher orders in $t$} 
  \label{app:higheramp}
For completeness, we here present the expansion of the tree-level four-point functions at the first few orders in $s$ and $t$. In the case of the open string expression \eqref{openamp} this expansion has the form (up to $O(t^3)$ and $O(s^5)$)   
\begin{align} \label{openexp}
& \tilde A_{op}(s,t) =   {\zeta}(2) +s  \zeta (3) +s^2  {\zeta}(4) +s^3  \zeta (5) +s^4  {\zeta}(6) +s^5  \zeta (7) +O (s^6)  \\ 
&  + t  \left( \zeta (3) +  \frac{{\zeta}(4)}{4}  s+ \left(2 \zeta (5)-{\zeta}(2) \zeta (3) \right) s^2+  \left( \frac{3 {\zeta}(6)}{4}-\frac{\zeta (3)^2}{2} \right) s^3 \right. \nn\\
&\left.  + \left( 3 \zeta (7) - {\zeta}(4) \zeta (3) - {\zeta}(2) \zeta (5) \right) s^4+ \left(  \frac{5 {\zeta}(8)} {4}- \zeta (3) \zeta (5) \right) s^5+O (s^6  )  \right) \nn \\
& +t^2  \left( {\zeta}(4) + \left( 2 \zeta (5)-{\zeta}(2) \zeta (3)\right)  s+  \left( \frac{23 {\zeta}(6)}{16}-  \zeta (3)^2\right)  s^2 \right.
\nn \\
& + \left(5 \zeta (7) - \frac{5 {\zeta}(4) \zeta (3)}{4} - 2 {\zeta}(2) \zeta (5) \right) s^3   + \left(\frac{61 \zeta(8)}{24}+ \frac{{\zeta}(2) \zeta (3)^2}{2}-3 \zeta (5) \zeta (3) \right)  s^4\nn\\
& +\left. \left( \frac{\zeta (3)^3}{6}-\frac{7}{4} {\zeta}(6) \zeta (3)-\frac{9}{4} {\zeta}(4) \zeta (5) -3 {\zeta}(2) \zeta (7)+\frac{28 \zeta (9)}{3} \right) s^5+O (s^6  )  \right)\nn\\
&+t^3 \left ( \zeta (5) +   \left( \frac{3 {\zeta}(6)}{4}-\frac{\zeta (3)^2}{2}\right)  s  + \left(5 \zeta (7) - \frac{5}{4} {\zeta}(4) \zeta (3)- 2 {\zeta}(2) \zeta (5)\right)  s^2 \right. \nn\\
& + \left( {\zeta}(2) \zeta (3)^2-4 \zeta (5) \zeta (3)+\frac{499 {\zeta}(8)}{192} \right) s^3 +  \left(\frac{\zeta (3)^3}{2}-\frac{35}{16} {\zeta}(6) \zeta (3)  -\frac{7}{2} {\zeta}(4) \zeta (5) -5 {\zeta}(2) \zeta (7) \right. 
\nn \\
& \left. +14 \zeta (9) \right) s^4 
+\left.  \left( \frac{9}{8} {\zeta}(4) \zeta (3)^2+3 {\zeta}(2) \zeta (5) \zeta (3)-8 \zeta (7) \zeta (3) -4 \zeta (5)^2  + {973 \zeta(10) \over 160}  \right)  s^5+O (s^6  ) \right )\, . \nn 
\end{align}

The expansion of the closed-string expression $ \tilde A^{(s)}_{cl}(s,t)$ up to $O(t^2)$ and $O(s^4)$  has the form 
\begin{align} \label{closedone}
& \tilde A^{(s)}_{cl}(s,t)  = \zeta (3) +s  {\zeta}(4) +s^2  \zeta (5) +s^3  {\zeta}(6) +s^4  \zeta (7) +O (s^5)  \\ 
&+2 t   \left( \frac{{\zeta}(4)}{4}  + \left(2 \zeta (5)-{\zeta}(2) \zeta (3) \right) s +  \left( \frac{3 {\zeta}(6)}{4}-\frac{\zeta (3)^2}{2} \right) s^2 \right. \nn \\
&\left.  + \left( 3 \zeta (7) - {\zeta}(4) \zeta (3) - {\zeta}(2) \zeta (5) \right) s^3+ \left(  \frac{5 {\zeta}(8)} {4}- \zeta (3) \zeta (5) \right) s^4+O (s^5  )  \right) \nn \\
&  +t^2  \left(  \frac{5\zeta(5)}{2} -  \zeta(2) \zeta(3) + \left({53  \zeta(6) \over 12}- 3 \zeta(3)^2\right)  s+ 
\left( 9\zeta(7) -3\zeta(2)\zeta(5) -3\zeta(3)\zeta(4) \right)  s^2 \right.  \nn\\
& + \left( \zeta(2,6) + {61 \zeta(8) \over 6}   +2 \zeta(2) \zeta(3)^2 - 12 \zeta(3) \zeta(5) \right) s^3 \nn\\
&  \left. + \left(  {113 \zeta(9)\over 6} -5\zeta(2)\zeta(7)-5\zeta(3)\zeta(6)-5\zeta(4)\zeta(5)+ {2\zeta(3)^3\over 3}  \right)  s^4 +O (s^5  )  \right) \,. \nn 
\end{align}
The low-energy expansion of $\tilde A^{(u)}_{cl}(u,t)$ is the same with $s$ and  $u=-t-s$ interchanged.  Each of these 
expressions contains even zeta values and irreducible MZVs -- for example, the coefficient of $t^2 s^3$  in  \eqref{closedone}  contains the weight-8 irreducible MZV $\zeta(2,6)$.   These cancel out in the low-energy expansion of the  total closed-string tree  amplitude, $\tilde A_{cl}(s,t) = \tilde A^{(s)}_{cl}(s,t)+\tilde A^{(u)}_{cl}(u,t)$, which has coefficients that are rational polynomials of odd zeta values.

\end{document}